\documentclass[twocolumn,english]{revtex4-1}
\usepackage[T1]{fontenc}
\usepackage[latin9]{inputenc}
\usepackage[a4paper]{geometry}
\usepackage{babel}
\usepackage{dcolumn}
\usepackage{bm} 
\usepackage{amsmath,amssymb,amsfonts}
\usepackage{verbatim}
\usepackage{graphicx,wrapfig,lipsum}
\usepackage{hyperref}
\usepackage{caption}
\usepackage{bbold}
\usepackage{float} 
\usepackage[usenames,dvipsnames]{xcolor}
\usepackage{epsfig}
\usepackage{epstopdf}
\usepackage{tikz}
\usetikzlibrary{shapes.geometric, arrows}
\usepackage{upgreek}
\usepackage{setspace}
\usepackage{enumitem}
\usepackage{array,multirow,bigdelim}

\def\dif{\text{d}}

\def\be{\begin{equation}}
\def\ee{\end{equation}}
\def\ba{\begin{eqnarray}}
\def\ea{\end{eqnarray}}

\def\a{\alpha}

\def\CP1{\mathbb{CP}^1}
\def\SL2C{\mathrm{SL}(2,\mathbb{C})}

\def\Z2{\mathbb{Z}_2}

\def\su2{{SU(2)}}

\def\a{{\alpha}}

\def\s{\sigma}
\def\a{\alpha}

\def\({\left(}
\def\){\right)}

\def\<{\langle}
\def\>{\rangle}

\def\i2{\frac{i}{2}}

\def\2F1{\,_2{\rm F}_1}

\begin{document}
\widetext

\title{Cosmological Scattering Equations}
\author{Humberto Gomez$^{a,b}$}
\author{Renann Lipinski Jusinskas$^{c}$}
\author{Arthur Lipstein$^{a}$}
\affiliation{$^{a}$Department of Mathematical Sciences, Durham University, Durham, DH1 3LE, UK}
\affiliation{$^{b}$Facultad de Ciencias Basicas,  Universidad Santiago de Cali,\\
Calle 5 $N^\circ$  62-00 Barrio Pampalinda, Cali, Valle, Colombia}

\affiliation{$^{c}$Institute of Physics of the Czech Academy of Sciences \& CEICO \\ 
Na Slovance 2, 18221 Prague, Czech Republic}

\begin{abstract}
We propose a worldsheet formula for tree-level correlation functions describing a scalar field with arbitrary mass and quartic self-interaction in de Sitter space, which is a simple model for inflationary cosmology. The correlation functions are located on the future boundary of the spacetime and are Fourier-transformed to momentum space. 
Our formula is supported on mass-deformed scattering equations involving conformal generators in momentum space and reduces to the CHY formula for $\phi^4$ amplitudes in the flat space limit. Using the global residue theorem, we verify that it reproduces the Witten diagram expansion at four and six points, and sketch the extension to $n$ points.
\end{abstract}

\maketitle

\section{Introduction}
Cosmological observations suggest the Universe underwent an early inflationary phase  approximately described by four-dimensional de Sitter (dS) space \cite{Guth:1980zm,Linde:1981mu,Albrecht:1982wi}. In this scenario, correlation functions on the dS future boundary are basic cosmological observables, encoding temperature fluctuations in the cosmic microwave background and the initial conditions for structure formation \cite{Mukhanov:1981xt}. These correlators can be computed using the in-in formalism \cite{Maldacena:2002vr,Weinberg:2005vy} or derived from the so-called wavefunction of the Universe \cite{Hartle:1983ai}, whose coefficients can be obtained by Wick rotating boundary correlators in Anti-de Sitter space (AdS) \cite{Maldacena:2002vr,McFadden:2009fg,McFadden:2010vh,Sleight:2020obc}. The wavefunction coefficients are constrained by conformal Ward identities (CWI) associated with the spacetime isometries, and can therefore be treated like correlation functions of a conformal field theory (CFT) living at the boundary \cite{Strominger:2001gp,Bzowski:2013sza,Bzowski:2015pba}. We will refer to them as cosmological correlators.

Perturbatively, cosmological correlators can be computed using Witten diagrams ending at the future boundary \cite{Liu:1998ty,Maldacena:2011nz,Raju:2011mp,Raju:2012zr,Ghosh:2014kba,Arkani-Hamed:2015bza}. In the flat space limit, they reduce to scattering amplitudes \cite{Raju:2012zr} for which a wealth of computational techniques has been developed, e.g. \cite{Britto:2005fq,Bern:1994cg,Arkani-Hamed:2016byb}. These methods have led to remarkable new formulations such as the Cachazo-He-Yuan (CHY) formulae \cite{Cachazo:2013hca,Cachazo:2013iea,Mason:2013sva}, recasting scattering amplitudes in a vast range of quantum field theories in terms of a universal set of scattering equations. This formulation has in turn manifested many remarkable structures \cite{Cachazo:2014xea} such as the double copy relating gauge and gravitational amplitudes \cite{Bern:2008qj,Bern:2010ue}.

By comparison, far less is known about cosmological correlators and the research programme for adapting flat space techniques to backgrounds with nonzero cosmological constant is still in its infancy 
\cite{Arkani-Hamed:2017fdk,Rastelli:2017udc,Caron-Huot:2017vep,Alday:2017vkk,Arkani-Hamed:2018kmz,Farrow:2018yni,Lipstein:2019mpu,Sleight:2019hfp,
Baumann:2019oyu,Armstrong:2020woi,Albayrak:2020fyp,Baumann:2020dch,Goodhew:2020hob,Bzowski:2020kfw,Meltzer:2020qbr,Melville:2021lst,Alday:2021odx,Jazayeri:2021fvk,Baumann:2021fxj,Zhou:2021gnu,Diwakar:2021juk}. 
Worldsheet formulas describing massless biadjoint scalars with cubic interactions in AdS were recently proposed in \cite{Eberhardt:2020ewh,Roehrig:2020kck}. Their scattering equations are written in terms of conformal generators acting on contact Witten diagrams in position space. 

In this letter, we propose a worldsheet formula for cosmological correlators describing scalar fields with arbitrary mass and quartic interaction in dS, which is one of the simplest models for inflation \cite{Martin:2013tda}. 
Our formula is based on a mathematical structure we call the cosmological scattering equations, which take a remarkably simple form in momentum space. They directly reduce to the CHY formula for $\phi^4$ amplitudes in the flat space limit \cite{Cachazo:2014xea}. Another nontrivial aspect of our construction is the presence of differential operators in the integrand in the form of a Pfaffian. Crucially, we find there are no ordering ambiguities. 

\section{Cosmological correlators}

We work in the Poincar\'e patch of $(d+1)$-dimensional dS with unit radius:
\begin{equation}
ds^{2}=\frac{-d\eta^{2}+(dx^i)^2}{\eta^{2}},
\end{equation}
where $-\infty<\eta<0$ is the conformal time, and $i=1,...,d$ runs over
Euclidean boundary directions. We will interchangeably use the notation $\vec{x}$ for boundary directions.

The $n$-point cosmological correlator, $\Psi_{n}$, can be treated as a CFT correlator in the future boundary, expressed in momentum space as
\begin{equation}
\Psi_{n}=\delta^{d}(\vec{k}_T)\langle \mathcal{O}(\vec{k}_{1})...\mathcal{O}(\vec{k}_{n})\rangle,
\label{psid}
\end{equation}
where $\vec{k}_T=\vec{k}_{1}+...+\vec{k}_{n}$. We will work with scalar operators $\mathcal{O}$ of scaling dimension $\Delta$,
dual to bulk scalar fields with mass
\begin{equation}
m^{2}=\Delta\left(d-\Delta\right).
\label{massdelta}
\end{equation}
The CWI for $\Psi_{n}$ can be expressed as
\begin{equation}
\sum_{a=1}^{n}P^i_a\Psi_{n}=\sum_{a=1}^{n}D_{a}\Psi_{n}=\sum_{a=1}^{n}K^i_a\Psi_{n}=0,
\label{cwi}
\end{equation}
where $a,b,...$ are particle labels and the conformal generators in momentum space are
\begin{eqnarray} \label{eq:boundary-gen}
P^{i} & = & k^{i}, \nonumber\\
D& = & k^{i}\partial_{i}+(d-\Delta), \label{eq:CGG-boundary}\\
K_{i} & = & k_{i}\partial^{j}\partial_{j}-2k^{j}\partial_{j}\partial_{i}-2(d-\Delta)\partial_{i}, \nonumber
\end{eqnarray}
with $\partial_i=\tfrac{\partial}{\partial k^i}$. 
Rotation generators act trivially on scalar operators so we do not include them. 

\section{Witten diagrams}

Cosmological correlators admit a perturbative expansion in terms of bulk Witten
diagrams ending on the future boundary. Here we take
the bulk theory to be a scalar with mass $m$ and quartic self-interaction.
The operators in the dual CFT have scaling dimension $\Delta$ satisfying \eqref{massdelta}.

The bulk-to-boundary propagator is
\begin{equation}
\mathcal{K}_{\nu}(k,\eta)=\mathcal{N}k^{\nu}\eta^{d/2}H_{\nu}(-k \eta),
\end{equation}
where $\nu=\Delta-d/2$, $k=|\vec{k}|$, $H_\nu$ is a Hankel function of the second kind, and $\mathcal{N}$ is a normalisation that we will not explicitly need. It satisfies
$ (\mathcal{D}_k^{2}+m^{2})\mathcal{K}_{\nu} =0$, with
\begin{equation}
\mathcal{D}_k^{2} \equiv \eta^{2}\partial_{\eta}^{2}+(1-d)\eta\partial_{\eta}+\eta^{2}k^{2}, \label{eq:D2def}
\end{equation}
and can be used to compute contact diagrams as follows:
\begin{eqnarray}
\mathcal{C}^{\Delta}_n & \equiv &  \int  \frac{d\eta}{\eta^{d+1}} U_{1,n}(\eta), \\
U_{1,n}(\eta) & = &  \prod_{a=1}^{n}\mathcal{K}_{\nu}(k_{a},\eta) \label{eq:prodK}.
\end{eqnarray}
As we will see, all tree-level Witten diagrams can be obtained from contact diagrams by acting with certain differential operators.

A central object in our analysis is the action of the operator
\begin{equation}\label{eq:10R}
\mathcal{D}_{a}\cdot\mathcal{D}_{b}= \tfrac{1}{2}(P_{a}^{i}K_{bi}+K_{ai}P_{b}^{i})+D_{a}D_{b},
\end{equation}
on the product $\mathcal{K}_{\nu}(k_{a},\eta)\mathcal{K}_{\nu}(k_{b},\eta)\equiv\mathcal{K}_{\nu}^{a}\mathcal{K}_{\nu}^{b}$. When acting on $\mathcal{K}_{\nu}$, the boundary generators in \eqref{eq:boundary-gen} can be written in terms of derivatives with respect to conformal time
\begin{eqnarray}
P^{i}\mathcal{K}_{\nu} & = & k^{i}\mathcal{K}_{\nu}, \nonumber\\
D\mathcal{K}_{\nu} & = & \eta\tfrac{\partial}{\partial\eta}\mathcal{K}_{\nu},  \\
K_{i}\mathcal{K}_{\nu} & = & \eta^{2}k_{i}\mathcal{K}_{\nu}, \nonumber
\end{eqnarray}
leading to
\begin{equation}
(\mathcal{D}_{a}\cdot\mathcal{D}_{b})\mathcal{K}_{\nu}^{a}\mathcal{K}_{\nu}^{b}=\eta^{2}[\partial_{\eta}\mathcal{K}_{\nu}^{a}\partial_{\eta}\mathcal{K}_{\nu}^{b}+(\vec{k}_{a}\cdot\vec{k}_{b})\mathcal{K}_{\nu}^{a}\mathcal{K}_{\nu}^{b}].\label{eq:DaDb}
\end{equation}
It is then straightforward to show that
\begin{align}
\mathcal{D}_{1...p}^{2}U_{1,n}&= (\mathcal{D}_{1}+\ldots + \mathcal{D}_{p})^2 U_{1,n}, \nonumber \\
& =  -p m^2 U_{1,n}+2\sum_{1\leq a<b\leq p}(\mathcal{D}_{a}\cdot\mathcal{D}_{b})U_{1,n}, \label{eq:bulkvsboundary-props}
\end{align}
where in the left hand side $\mathcal{D}_{1...p}^{2}$ is defined in \eqref{eq:D2def} with $k=|\vec{k}_{1}+...+\vec{k}_{p}| \equiv k_{1...p}$ and $p<n$,
and the right hand side is built using the boundary conformal generators
in momentum space (\ref{eq:CGG-boundary}), satisfying $\mathcal{D}_a \cdot \mathcal{D}_a  = -m^2$.

In practice we will encounter the inverse of boundary differential operators constructed from those in \eqref{eq:10R}. Using \eqref{eq:bulkvsboundary-props}, we replace them with the inverse of the bulk differential operator in \eqref{eq:D2def} leading to bulk-to-bulk propagator insertions:
\begin{multline}
[(\mathcal{D}_1+\ldots +\mathcal{D}_p)^2+m^2]^{-1} \mathcal{C}^{\Delta}_n = \\  \int  \frac{d\eta}{\eta^{d+1}}  \frac{d\tilde{\eta}}{\tilde{\eta}^{d+1}}   U_{p+1,n}(\eta) G_{\nu}(k_{1...p},\eta,\tilde{\eta})U_{1,p} (\tilde{\eta}),
\end{multline}
with
$(\mathcal{D}_{k}^{2}+m^{2})G_{\nu}(k,\eta,\tilde{\eta})=\eta^{d+1}\delta(\eta-\tilde{\eta})$.

\section{Scattering equations}

In flat space, the CHY formulae express tree-level scattering amplitudes as integrals over the Riemann sphere, mapping each external leg to a puncture. The integrals then localise onto solutions of the scattering equations (SE):
\begin{equation}
\sum_{a\neq b}\frac{2\, k_{a}\cdot k_b}{\sigma_{ab}}=0, \qquad \sigma_{ab} \equiv \sigma_{a}-\sigma_{b},
\end{equation}
where $\sigma_a$ is the holomorphic coordinate of the $a$'th puncture. 
Inspired by the massive scattering equations of \cite{Dolan:2013isa} and the ambitwistor string formulae in AdS \cite{Eberhardt:2020ewh,Roehrig:2020kck}, we define the scattering equations in dS momentum space in terms of the following differential operators:
\begin{align}\label{eq:20R}
S_a = \sum_{b=1 \atop b\neq a}^{n} \frac{2\,  ({\cal D}_a \cdot {\cal D}_b )+\mu_{ab} }{\sigma_{ab}} \equiv \sum_{b=1 \atop b\neq a}^{n} \frac{\a_{ab} }{\sigma_{ab}} \, \, ,
\end{align}
where  $\mu_{a\,a\pm 1} =-m^2$ modulo $n$ and zero otherwise. This mass deformation assumes canonical ordering of the external legs $\mathbb{1}_n = (1,2,\ldots ,n)$. Different orderings are obtained by permutations. 

Using   the CWI in \eqref{cwi}, we can show that 
\begin{equation}
\sum_{a\neq b}\alpha_{ab}=0,
\end{equation}
which implies that the SE have an underlying ${\rm SL}(2,\mathbb{C})$ symmetry \cite{W.in.P}. It can then be used to fix the location of three punctures using a standard procedure familiar from string theory.

A generic worldsheet integral will then take the form
\begin{equation}\label{eq:22R}
\int_\gamma  \prod_{a=1 \atop a\neq b,c,d}^n  \dif\sigma_a \, (S_a)^{-1} \,  (\sigma_{bc}\sigma_{cd}\sigma_{db})^2 \, \mathcal{I}_n,
\end{equation}
where the integration contour is defined by the intersection $\gamma= \bigcap_{a\neq b,c,d} \gamma_{S_a}$, where $\gamma_{S_a}$ encircles the pole where $S_a$ vanishes when acting on the theory-dependent integrand $\mathcal{I}_n$.
Following similar steps to \cite{Eberhardt:2020ewh,Roehrig:2020kck}, it is possible to show that the differential operators in \eqref{eq:20R} commute, so the measure in \eqref{eq:22R} is well-defined.  

\section{Worldsheet Formula}
Using the cosmological SE defined in the previous section, we now propose a worldsheet formula for cosmological correlators describing massive $\phi^4$ theory in dS momentum space: 
\begin{equation}\label{eq:totalphi4}
\Psi_n=  \frac{ \delta^{d}(\vec{k}_T)}{(3!)^{p-1}} \!\! \sum_{\rho\in {\rm S}_{n-1}} \!\!\!{\rm sgn}_\rho\,\,  {\cal A}(\rho(1,2,\ldots,n-1),n) \, {\cal C}^{\Delta}_n,
\end{equation}
where $n=2p \in \rm{even}$, ${\rm S}_{n-1}$ is the permutation group and
\begin{equation}\label{eq:7}
{\cal A}(\mathbb{1}_n ) \!= \int_\gamma  \prod^n_{a\neq b,c,d}  \dif \sigma_a\, S_a^{-1}   (\s_{bc} \s_{cd} \s_{db} )^2  \,\,  {\cal I}(\mathbb{1}_n) ,
\end{equation}
with
\begin{align}\label{}
{\cal I}(\mathbb{1}_n)= 
 {\rm PT}(\mathbb{1}_n) \,\,\,
{\rm Pf}^\prime A\, \times
\!\!\!\!\!\!\!\!\!\!
 \sum_{\{a,b\}\in cp(\mathbb{1}_n)} \frac{{\rm sgn}(\{a,b\})  }{  \s_{a_1b_1}\cdots \s_{a_{p} b_{p} } }   . 
\end{align}
Here  ${\rm PT}(\mathbb{1}_n) = \left(\sigma_{12}\sigma_{23}...\sigma_{n1}\right)^{-1}$, $cp(\mathbb{1}_n)$ denotes all perfect matchings that lead to connected graphs related to the ordering $(1,2,...,n)$ \cite{Cachazo:2014xea}, and the reduced Pfaffian ${\rm Pf}^\prime A$ is given by 
\begin{equation}
{\rm Pf}^\prime A=\frac{(-1)^{c+d}}{\sigma_{cd}}  {\rm Pf}A^{cd}_{cd},
\end{equation}
where
\begin{align}
{\rm Pf}A^{cd}_{cd}=
\frac{ \epsilon^{r_1 s_1\ldots r_{p-1} s_{p-1}} (A^{cd}_{cd})_{r_1 s_1} \cdots (A^{cd}_{cd})_{r_{p-1} s_{p-1}}   }{2^{p-1}(p-1)!}.
\end{align}
The matrix $A^{cd}_{cd}$ is obtained from the $n \times n$ matrix
\begin{align}
A_{rs} & =
\begin{cases} 
\displaystyle \frac{ \a_{rs} }{\sigma_{rs}},  & r\neq s,\\
\displaystyle  ~~ 0 , & r=s,
\end{cases}
\end{align}
by removing any pair of rows and columns $\{c,d\}$. 

Since $[\a_{rs}, \a_{pq}]=0$ for $r\neq s\neq p \neq q$, and
$\sum_a [ P_a^i,\a_{rs} ]=\sum_a [ D_a,\a_{rs} ]= \sum_a [ K_a^i,\a_{rs} ]=0$,
${\rm Pf}^\prime A$ is well defined and $\Psi_n$ satisfies the CWI. 

\section{Flat Space Limit}

As a first test of our formula, let us check the flat space limit $E\rightarrow 0$, where $E = k_1+...+k_n$.
This limit can be accessed by taking $\eta\rightarrow-\infty$
in the integrand of the correlator \cite{Raju:2012zr,Lipstein:2019mpu}. Using the asymptotic form of the
bulk-to-boundary propagators,
\begin{equation}
\lim_{\eta\rightarrow-\infty}\mathcal{K}_{\nu}(k,\eta)\propto k_{i}^{\nu-1/2}\eta^{(d-1)/2}e^{ik\eta},
\end{equation}
equation \eqref{eq:DaDb} leads to
\begin{equation}
\lim_{\eta\rightarrow-\infty}\mathcal{D}_{a}\cdot\mathcal{D}_{b}\left(\mathcal{K}_{\nu}^{a}\mathcal{K}_{\nu}^{b}\right)=\eta^{2}(k_{a}\cdot k_{b})\mathcal{K}_{\nu}^{a}\mathcal{K}_{\nu}^{b},
\end{equation}
with $(k_{a}\cdot k_{b})= -k_a k_b+\vec{k}_a\cdot \vec{k}_b$. In this limit we can therefore replace $\mathcal{D}_{a}\cdot\mathcal{D}_{b}$
with $k_{a}\cdot k_{b}$ and set $m=0$ (recall that the mass is defined in units of the inverse dS radius so in the flat space limit it will vanish). 

The resulting conformal time integration then gives
\begin{equation}
\lim_{E\rightarrow0}\Psi_{n}\propto E^{-\left(4-d+\frac{1}{2}(d-3)n\right)}\Pi_{i=1}^{n}k_{i}^{\nu-1/2}\mathcal{A}_{n}\delta^{d}( \vec{k}_T),
\end{equation}
where $\mathcal{A}_{n}$ is the CHY formula for massless $\phi^{4}$
amplitudes in flat space. We have only kept contributions which arise from acting with differential operators directly on bulk-to-boundary propagators, since other contributions are subleading.

\section{Four points}

\begin{figure}
\centering
(a)
\!\!\!\!\!\!\!\!\!\!\!\!\!\!\!\!\!
\parbox[c]{12.2em}{\includegraphics[scale=0.23]{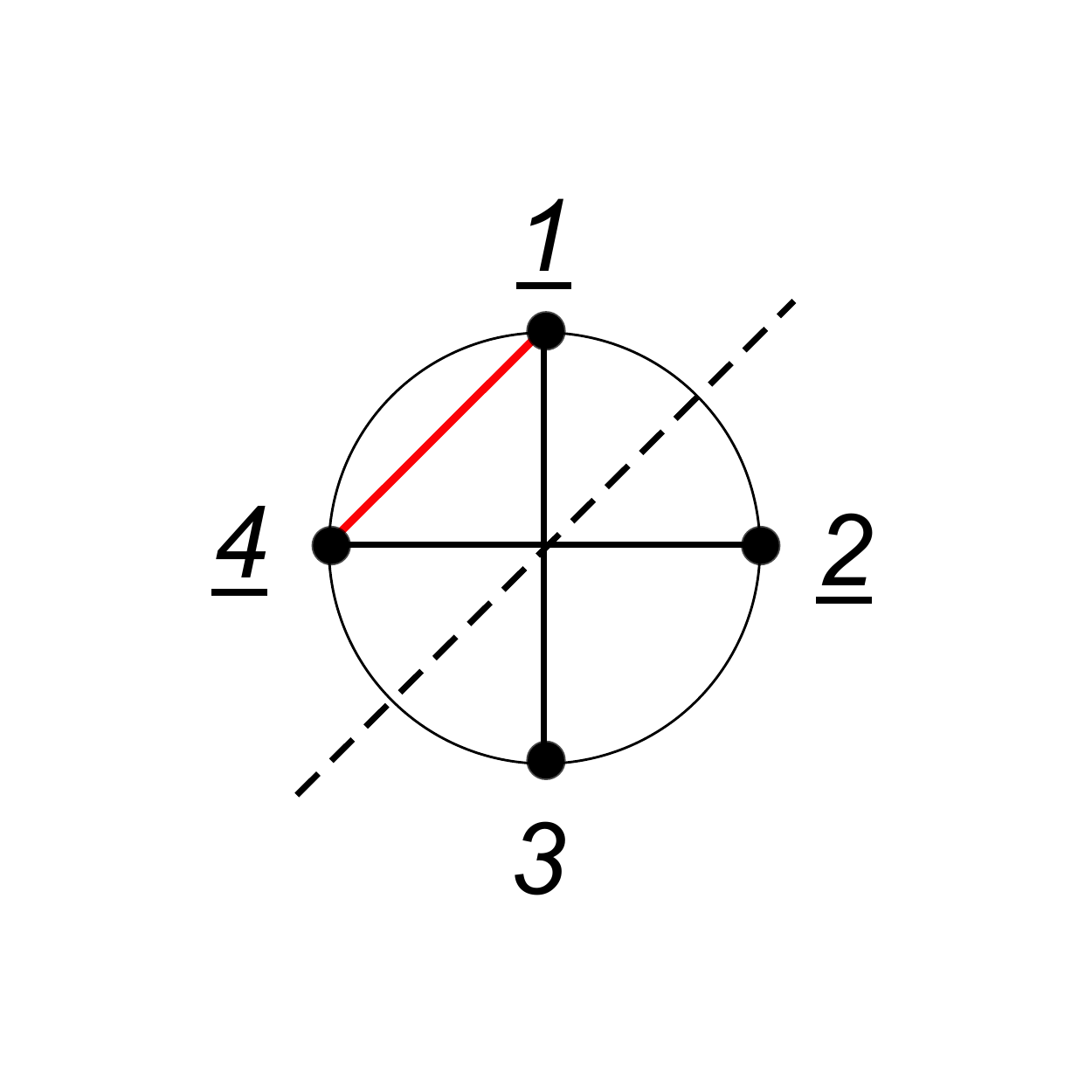}}
(b)
\!\!\!\!\!\!
\parbox[c]{8.5em}{\includegraphics[scale=0.23]{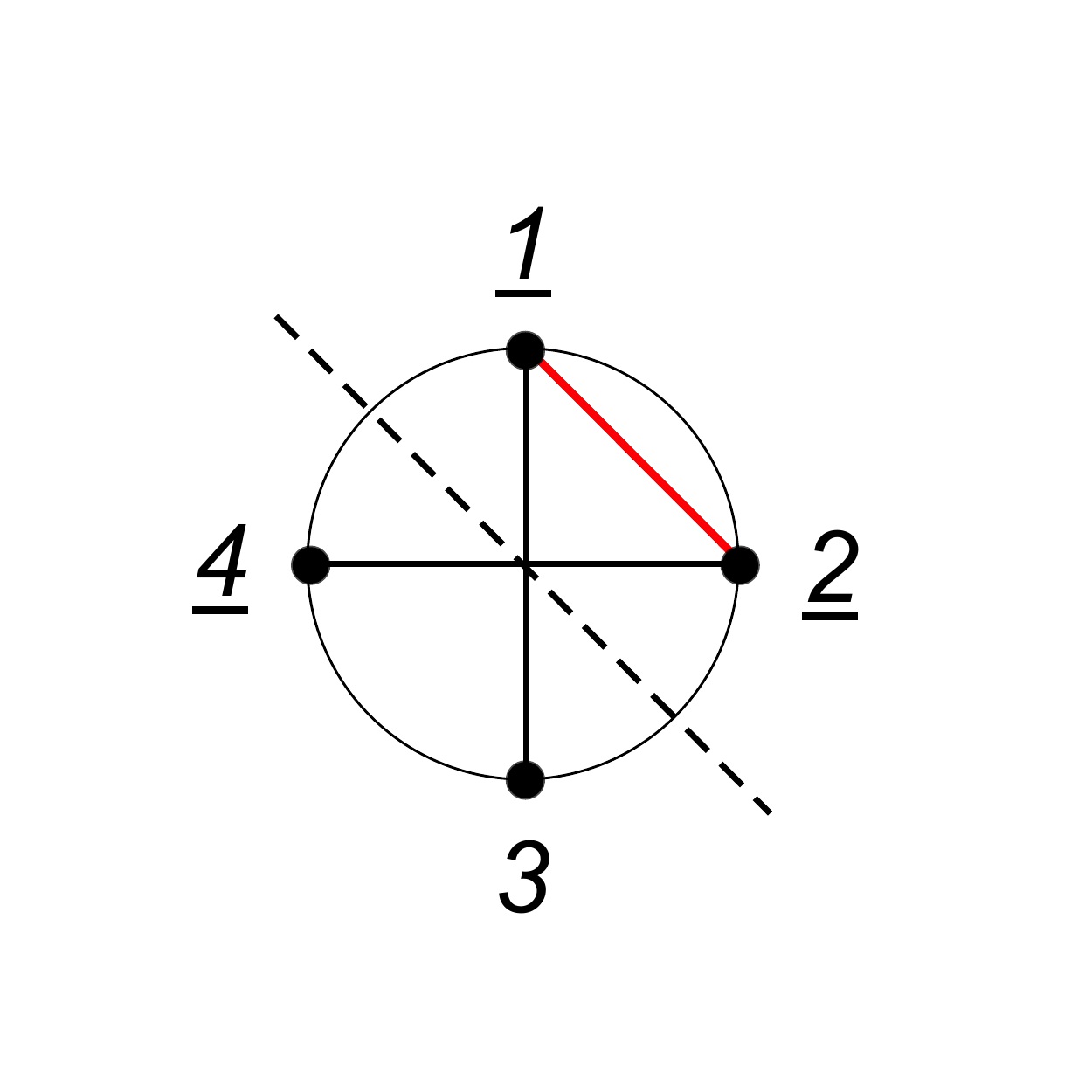}} 
\vspace{-0.45cm}
\caption{Factorization of ${\cal A}(\mathbb{1}_4)$ with (a) ${\rm Pf} A^{14}_{14}$, (b) ${\rm Pf} A^{12}_{12}$.}\label{Fig2} 
\vspace{-0.6cm}
\end{figure}

Let us first consider the ordered correlator
\begin{equation}\label{eq:11}
{\cal A}(\mathbb{1}_4){\cal C}^{\Delta}_4 \!= \!\! \int_{\gamma} \! \dif\sigma_3 ( \sigma_{41} \sigma_{12} \sigma_{24} )^2 S_3^{-1}  
\frac{   {\rm PT}(\mathbb{1}_4) (-1)  {\rm Pf} A^{14}_{14} }{\sigma_{14} \, (\sigma_{13} \sigma_{24})}  {\cal C}^{\Delta}_4 ,
\end{equation}
Notice that ${\cal A}(\mathbb{1}_4)$ has a graph representation, given in Fig. \ref{Fig2}.
The circle is the Parke-Taylor factor, ${\rm PT}(\mathbb{1}_4)$, the black lines depict the perfect matching and the red line indicates the rows/columns removed from the $A$-matrix. The underlined labels $\{4,1,2\}$ are the coordinates fixed by the ${\rm SL}(2,\mathbb{C})$ symmetry. Writing $\tilde S_a =S_a\times\prod_{b\neq a}^4 (\sigma_{ab})$ we obtain
\begin{align}\label{eq:13}
\!\!
{\cal A}(\mathbb{1}_4)\, {\cal C}^{\Delta}_4= -  \int_{\tilde\gamma}\dif\sigma_3 \, \tilde S_3^{-1}\,\,  
\frac{  \sigma_{12}  \, \sigma_{24}  }{\sigma_{32} } \, \a_{23}\, {\cal C}^{\Delta}_4 \, ,
\end{align}
where $\tilde\gamma$ contour is defined by $\tilde S_3$. Using the global residue theorem (GRT) \cite{Harris}, $\tilde\gamma$ can be deformed to $\gamma_{32}$, with $\gamma_{ab}=\{|\sigma_a-\sigma_b|=\epsilon\}$. Noting that $\tilde{S}_{3}|_{\sigma_{32}=0}= \sigma_{12}  \sigma_{42} \alpha_{23}$ and integrating around $\gamma_{32}$ then gives
\begin{align}
\!\!
{\cal A}(\mathbb{1}_4)  \,  {\cal C}_4^\Delta =  
( \a_{23} )^{-1}
(\a_{23})  \, \,   {\cal C}_4^\Delta ={\cal C}_4^\Delta , 
\end{align}
which is the desired result. Switching the order of the Pfaffian and the SE in \eqref{eq:11} leads to the same expression.

When $\sigma_3\rightarrow \sigma_2$, the worldsheet factorizes into two spheres. This can be visualised by cutting the planar graph with a dotted line as shown in Fig. \ref{Fig2}(a). On the other hand the factorizations $\sigma_3\rightarrow \sigma_1$ and $\sigma_3\rightarrow \sigma_4$ do not contribute. These observations motivate the following rules \cite{W.in.P}:

1) If all fixed points (underlined labels) are on the same side of a cut then this contribution vanishes because after factorization the two new spheres must each have three fixed punctures as shown in Fig \ref{Fig5}. 

2) If a factorization cuts more than four lines in the corresponding planar graph then this contribution vanishes. 
For example, in Fig. \ref{Fig2}(b) the only contribution is given by the factorization $\sigma_3 \rightarrow \sigma_4$.

Because of the ${\rm SL}(2,\mathbb{C})$ symmetry, the correlator is independent of the choice of fixed punctures. These rules help to identify the most convenient choices for computations. In \cite{W.in.P}, we will explicitly show that different choices are equivalent.

\section{Six points}
\begin{figure}
\centering
\hspace{-0.7cm}
\parbox[c]{5.1em}{\includegraphics[scale=0.2]{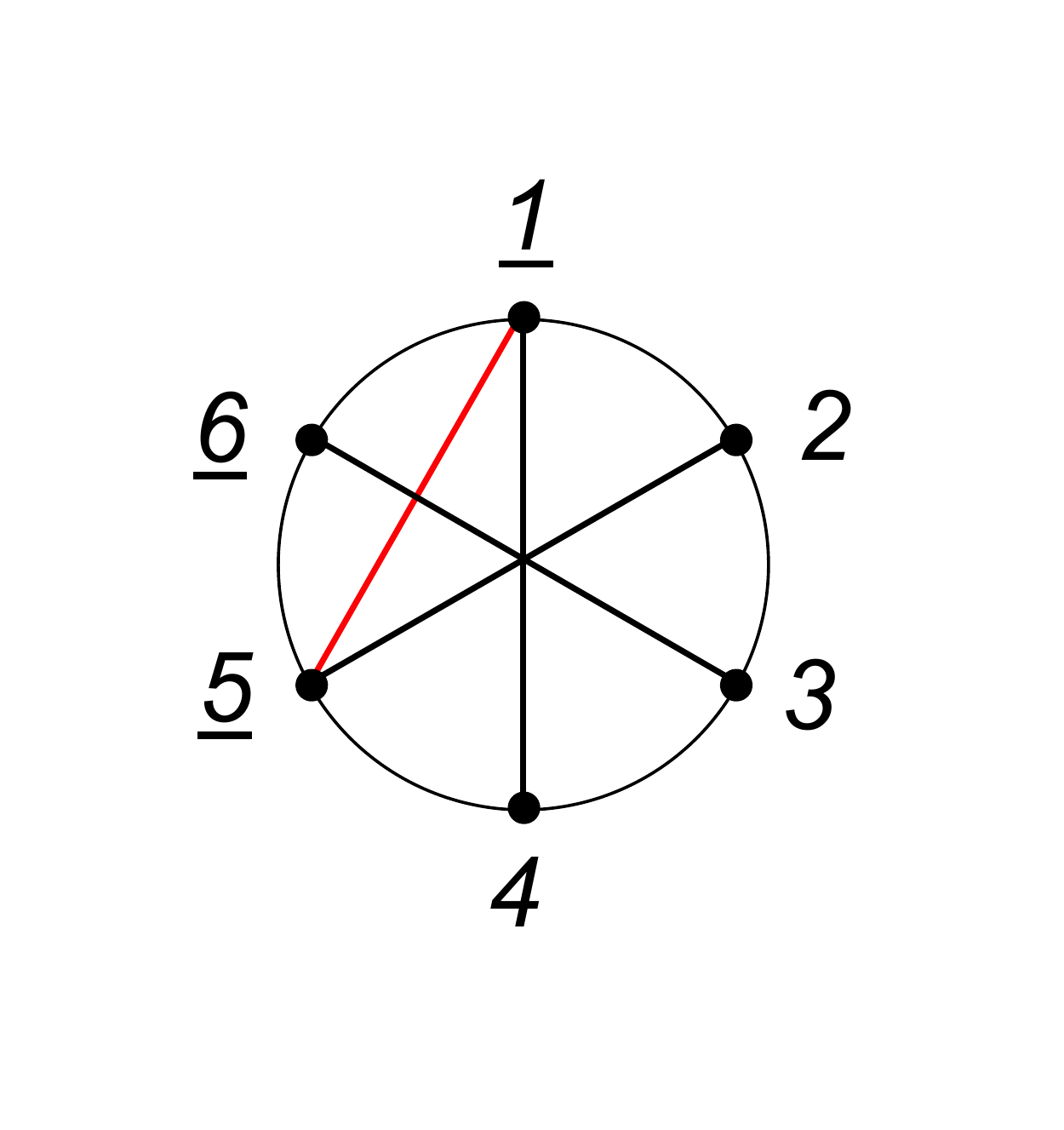}}\,\, \,
\parbox[c]{5.1em}{\includegraphics[scale=0.2]{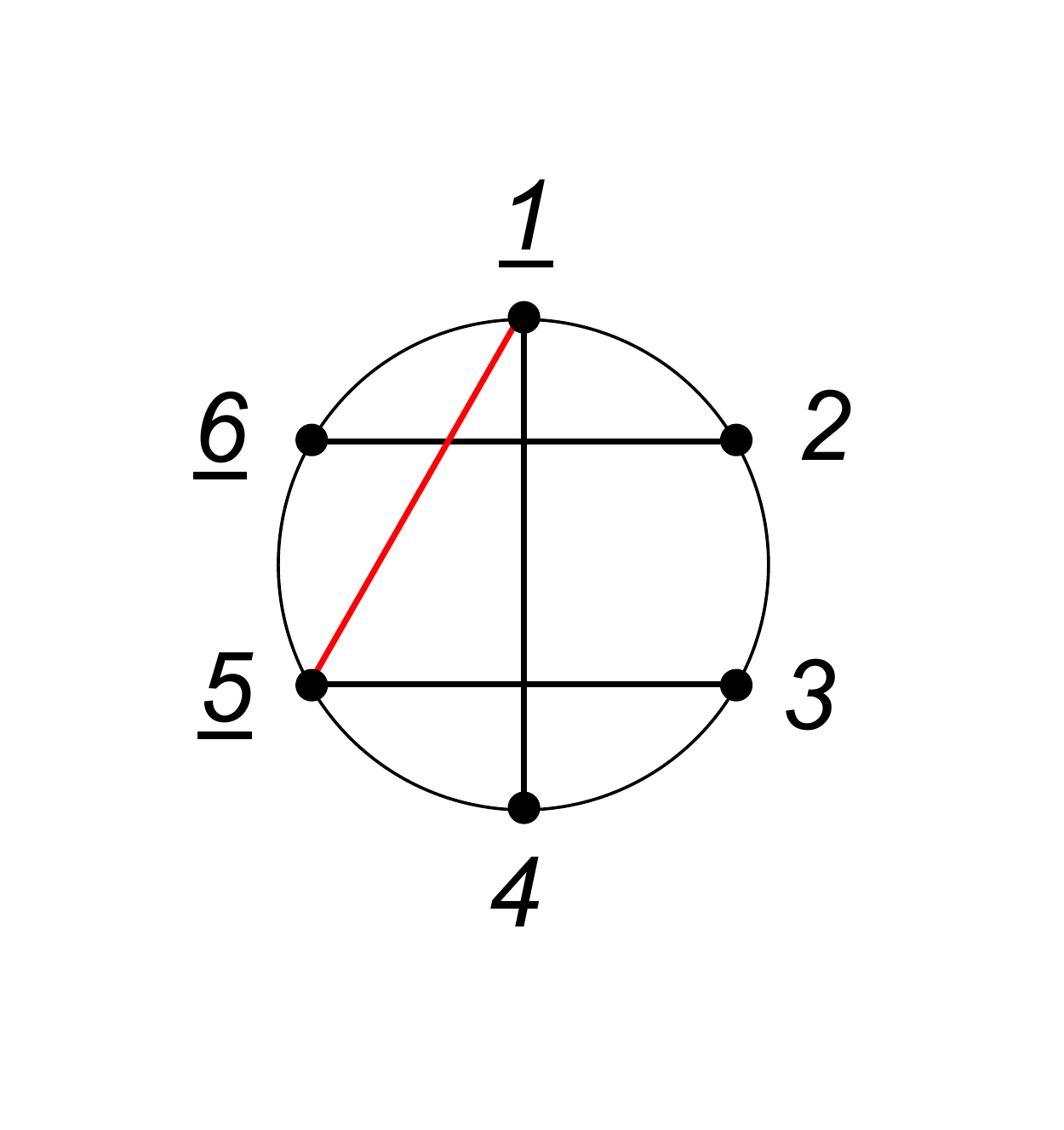}} \,  \,\, 
\parbox[c]{5.1em}{\includegraphics[scale=0.2]{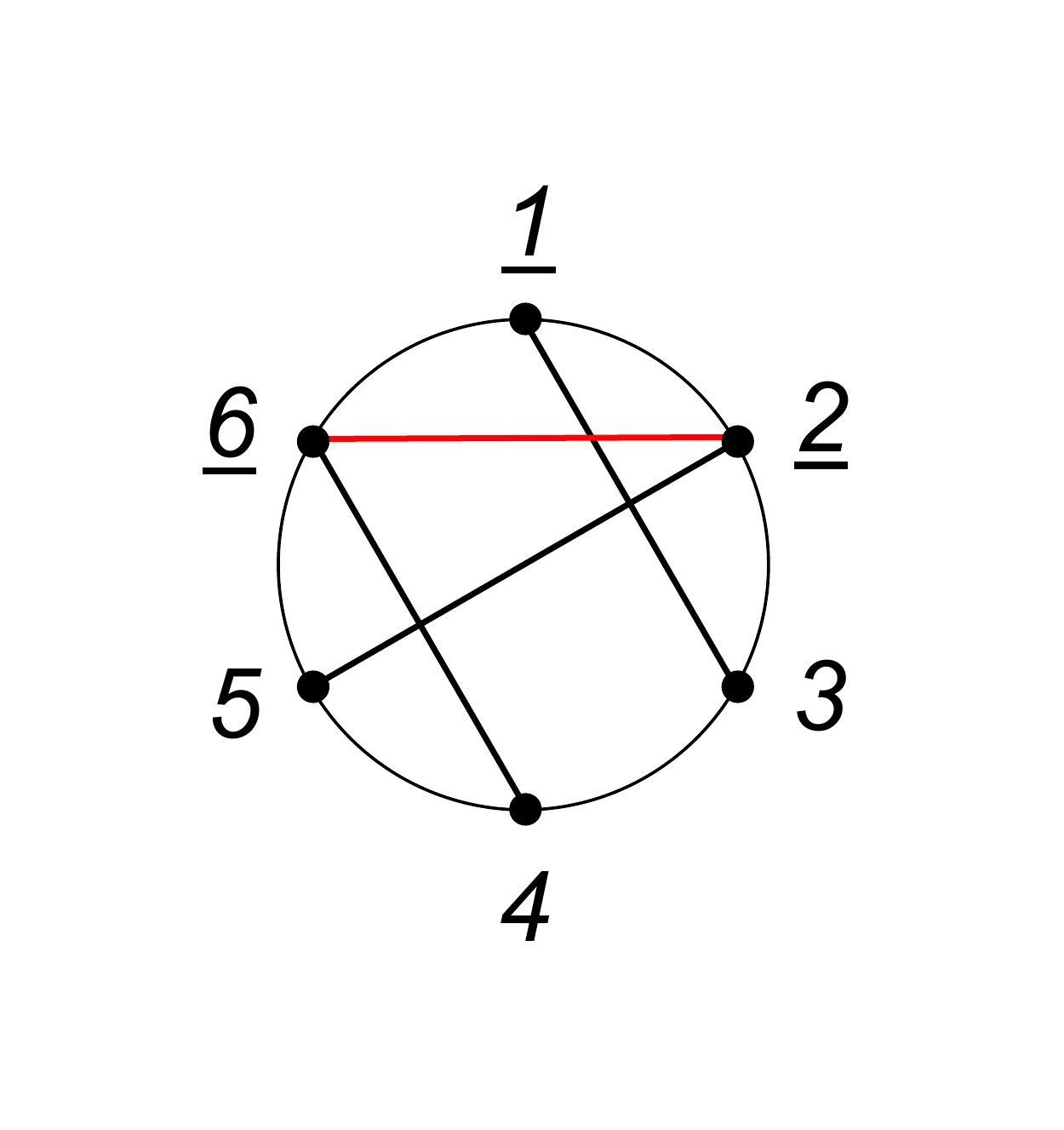}} \, \,\, 
\parbox[c]{5.1em}{\includegraphics[scale=0.2]{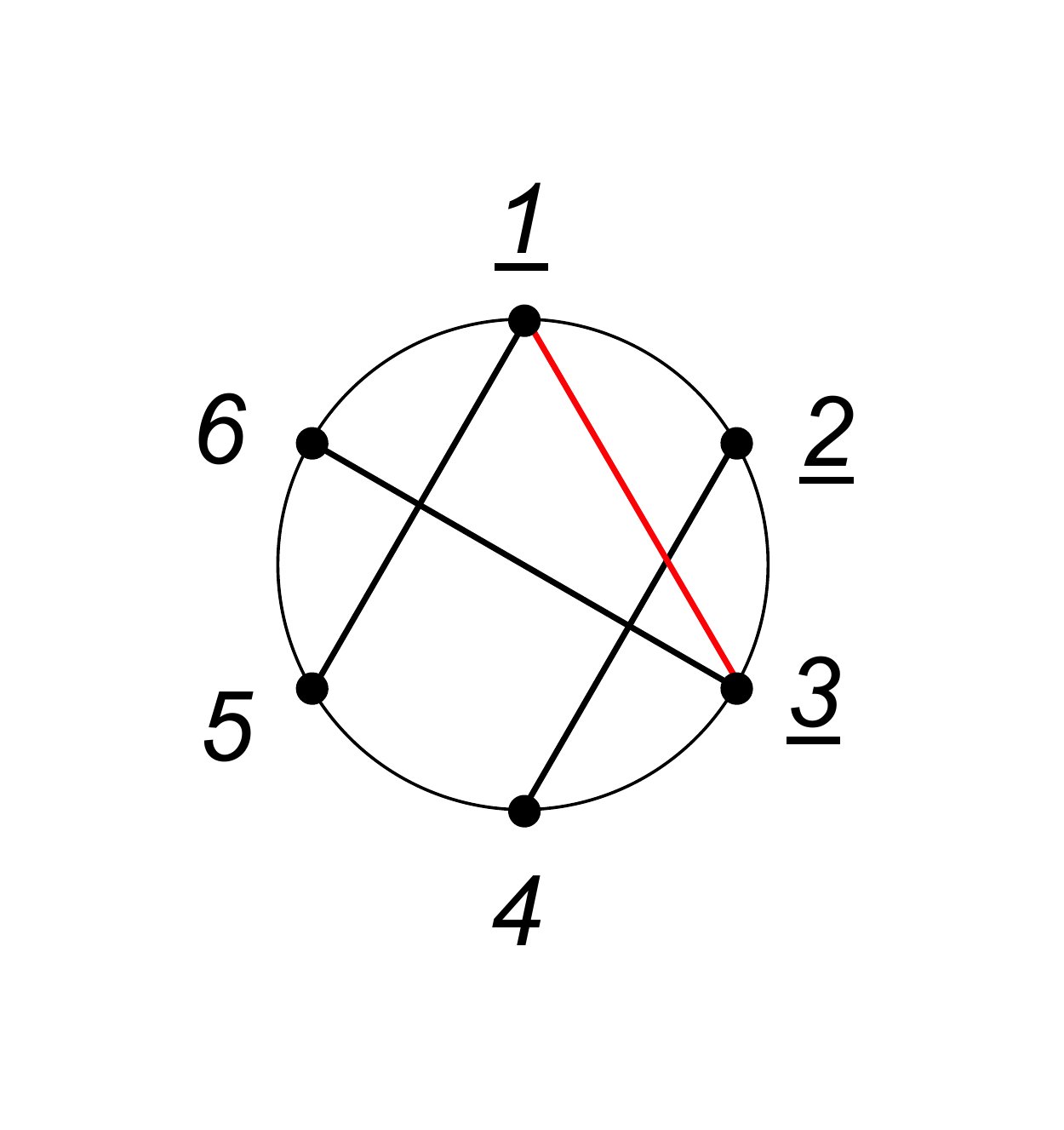}}   
\vspace{-0.45cm}
\caption{Diagrammatic representation of ${\cal A}(\mathbb{1}_6)$.}\label{Fig4} 
\vspace{-0.6cm}
\end{figure}

From \eqref{eq:7}, we see that ${\cal A}(\mathbb{1}_6)$ 
is encoded by the four diagrams in Fig. \ref{Fig4}.
In the first and second diagrams, we have fixed legs $\{5,6,1\} $ and removed the rows/columns $\{ 1,5\}$ from the $A$-matrix in the reduced Pfaffian. This is the simplest option. Other Pfaffian choices lead to additional contributions from the contour integrals which cancel out.

\begin{figure}[h]
\centering
\parbox[c]{8.2em}{\includegraphics[scale=0.23]{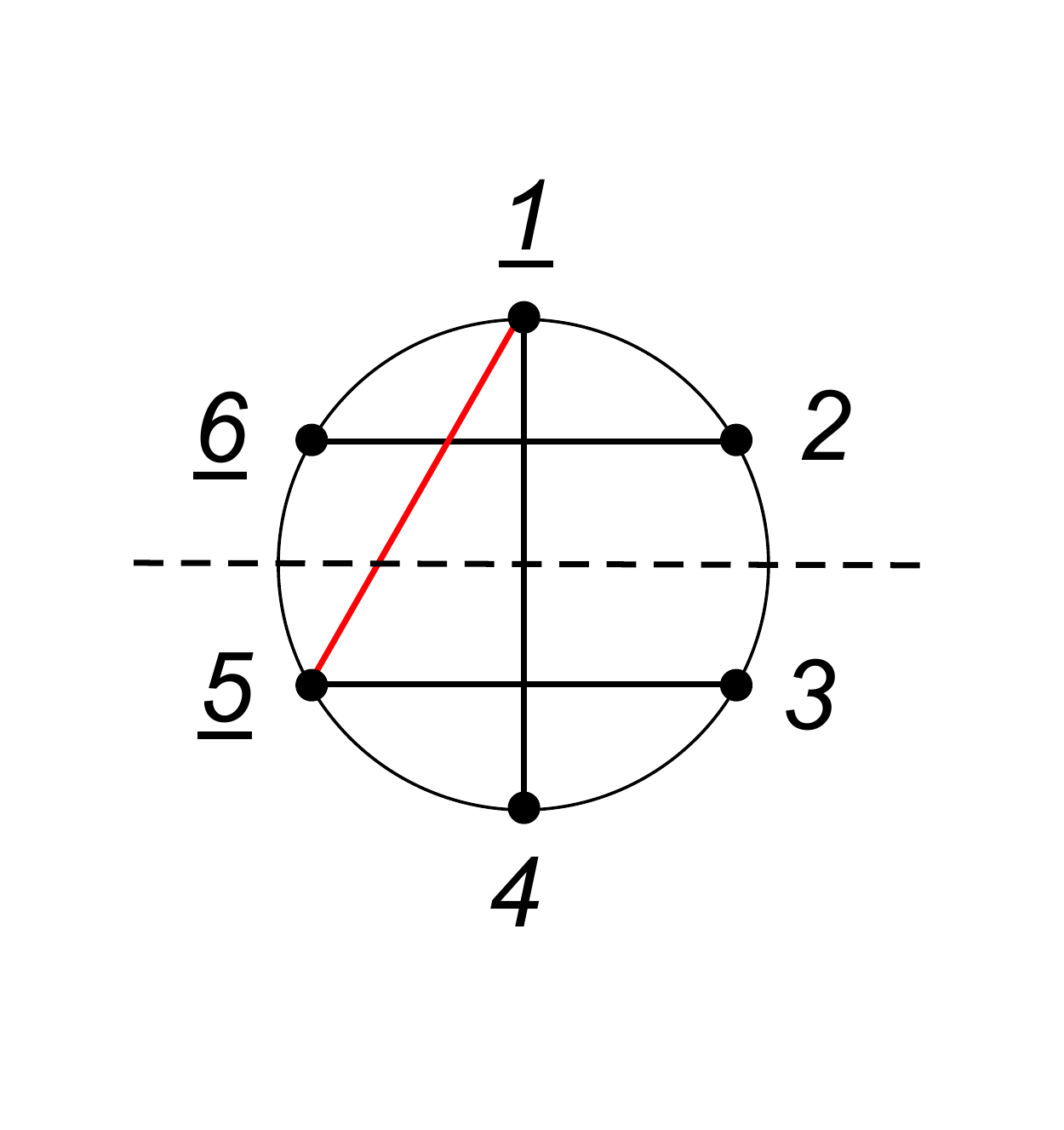}}
$\Rightarrow$
\parbox[c]{10.2em}{\includegraphics[scale=0.11]{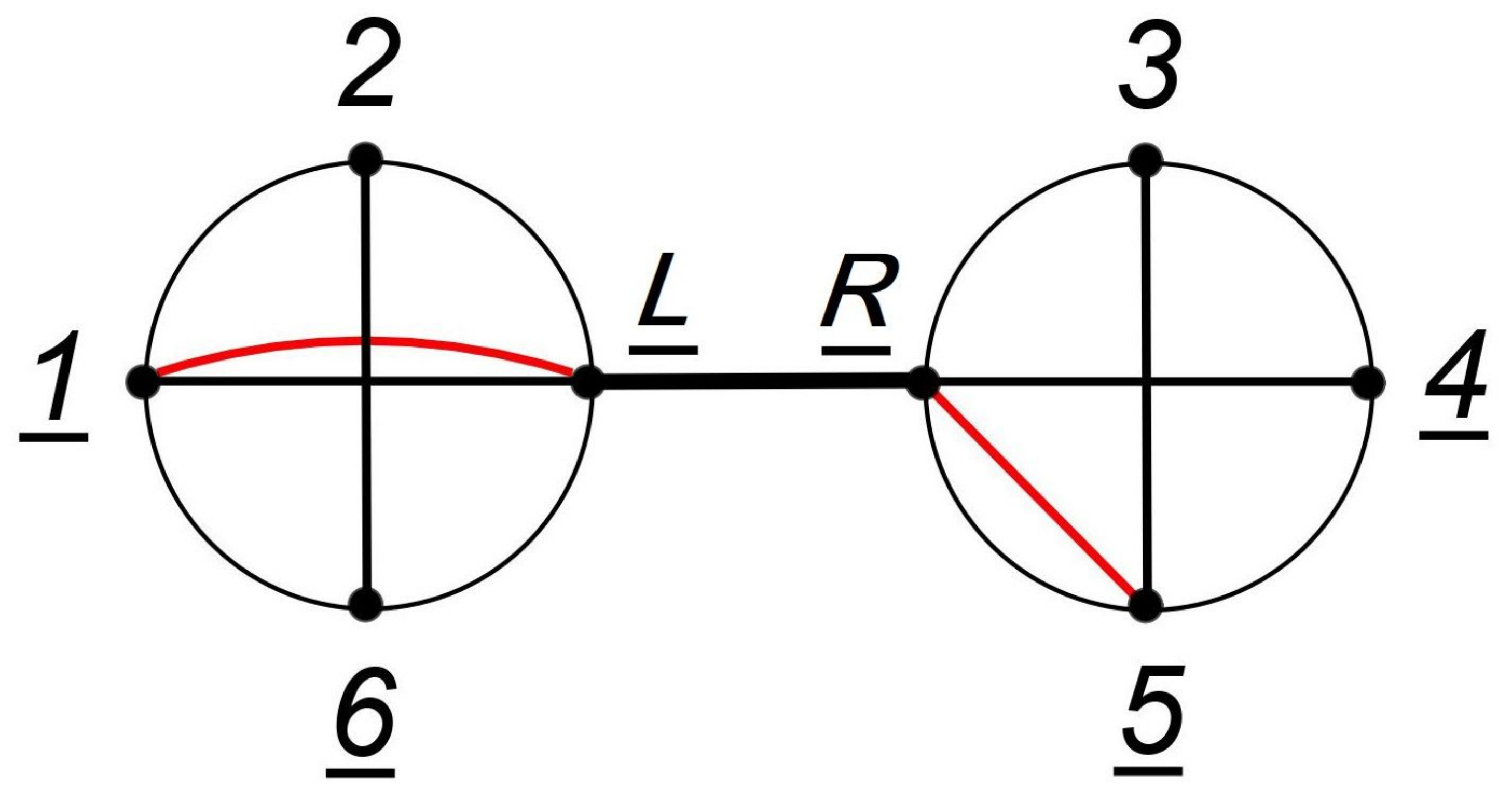}} 
\vspace{-0.45cm}
\caption{Factorization contribution.}\label{Fig5} 
\vspace{-0.4cm}
\end{figure}

Rules 1 and 2 tell us the first diagram has only one factorization contribution,  $\sigma_3\rightarrow\sigma_6$. After using the GRT we find that this diagram vanishes. 
The last three diagrams are identical up to cyclic permutations. We focus on the second one, {\it i.e.} ${\cal A}(\mathbb{1}_6:14,26,35) $, where the second argument in $ {\cal A} $ denotes the perfect matching.
Rules 1 and 2 imply two factorizations: $\sigma_2\rightarrow \sigma_6$ and $\sigma_3 \rightarrow\sigma_4\rightarrow \sigma_5$. Moreover, we find that the factorization $\sigma_2\rightarrow \sigma_6$ vanishes, so the only contribution comes from the latter (Fig. \ref{Fig5}). To compute it, we consider the parametrization $\sigma_a=\epsilon x_a+\sigma_5$,  
with $a=3,4,5$, $x_4=\textrm{constant}$, $x_5=0$, $\sigma_5\equiv\sigma_L$, and expand around $\epsilon=0$. The SE reduce to
\vspace{-0.5cm}
\vspace{-0.0cm}
\noindent
\begin{align}\label{eq:13}
& S_2 = [\hat S_2 + {\cal O}(\epsilon) ] , ~ \hat S_2=\frac{\a_{21}}{\sigma_{21}}  +\frac{\a_{26}}{\sigma_{26}} + \frac{\a_{2L}}{\sigma_{2L}}, \nonumber\\
& S_3 = \frac{1}{\epsilon}   [ \hat S_3  + {\cal O}(\epsilon) ]  ,  ~ \hat S_3=\frac{\a_{34}}{x_{34}}  +\frac{\a_{35}}{x_{35}} + \frac{\a_{3R}}{x_{3R}} , \\
& S_4 = \frac{1}{\epsilon}   [ \hat S_4 + {\cal O}(\epsilon) ]  , ~ \hat S_4=\frac{\a_{43}}{x_{43}}  +\frac{\a_{45}}{x_{45}} + \frac{\a_{4R}}{x_{4R}} \nonumber  ,
\end{align}
where $x_{R}=\infty$, $\a_{2L}= \a_{23}+\a_{24}+\a_{25}$ and $\a_{a R}= \a_{a6}+\a_{a1}+\a_{a2}$, $a=3,4$. Using the GRT, the contour can then be deformed to $\hat\gamma= \gamma_{\epsilon}\cap\gamma_{ \hat{S}_2}\cap\gamma_{ \hat{S}_3}$, with $\gamma_{\epsilon} = \{ |\epsilon|=\delta \}$. After performing the integral over $\epsilon$ and noting that $\hat{S}_{4}|_{\gamma_{\hat{S}_{3}}}=\frac{x_{R5}}{x_{54}\,x_{4R}}[({\cal D}_{6}+{\cal D}_{1}+{\cal D}_{2})^{2}+m^{2})]$, the remaining contour integral factorizes according to Fig. \ref{Fig5}:
\begin{align}\label{eq:14}
&
\!\!
{\cal A}(\mathbb{1}_6:14,26,35) \, \, {\cal C}_6^\Delta = {\cal A}(6,1,2,L:1L,26)   \nonumber  \\
&
\!\!
[ ( {\cal D}_{3}+{\cal D}_{4}+{\cal D}_{5})^2+m^2 ]^{-1}
{\cal A}(R,3,4,5:R4,35) \, \, {\cal C}_6^\Delta , 
\end{align}
\vspace{-0.3cm}
\noindent
where 
\begin{align}\label{eq:15}
&
\!\!
{\cal A}(6,1,2,L:1L,26) = \nonumber\\
&
\!\!
\int_{\gamma_{\hat S_2}} \!\!\! \dif\sigma_2 (\sigma_{1L} \sigma_{L 6} \sigma_{61})^2 \hat{S}_2^{-1}  {\rm PT}(6,1,2,L) \,
    \frac{{\rm Pf}A^{1L}_{1L}}{(\sigma_{1L}\sigma_{26}) \sigma_{1L}}   ,
\end{align}
\vspace{-0.5cm}
\noindent
\begin{align}\label{eq:16}
&
\!\!
{\cal A}(R,3,4,5:R4,35)  = \nonumber\\
&
\!\!
\int_{\gamma_{\hat S_3}} \!\!\!\dif x_3 (x_{45} x_{5R} x_{R4})^2 \hat{S}_3^{-1} \, {\rm PT}(R,3,4,5) \,
 \frac{(-1) {\rm Pf}A^{R5}_{R5}}{ (x_{R4}x_{35}) x_{R5}}  . 
\end{align}
Finally, using the result of the previous section we obtain
\begin{equation}\label{eq:17}
{\cal A}(\mathbb{1}_6:14,26,35) {\cal C}_6^\Delta \!= \!  [ ( {\cal D}_{3}+{\cal D}_{4}+{\cal D}_{5})^2+m^2 ]^{-1}  {\cal C}_6^\Delta,
\end{equation}
which is the Witten diagram for two 4-point vertices connected by a bulk-to-bulk propagator. Since all terms in \eqref{eq:14} commute, and the scattering equations commute with Pfaffians in each four-point integrand, this implies that shuffling terms in the Pfaffian with the scattering equations in the original expression leaves the final result unchanged.

\section{$n$ points}
\begin{figure}
\quad
(a)
\!\!\!\!\!\!\!
\parbox[c]{11.2em}{\includegraphics[scale=0.22]{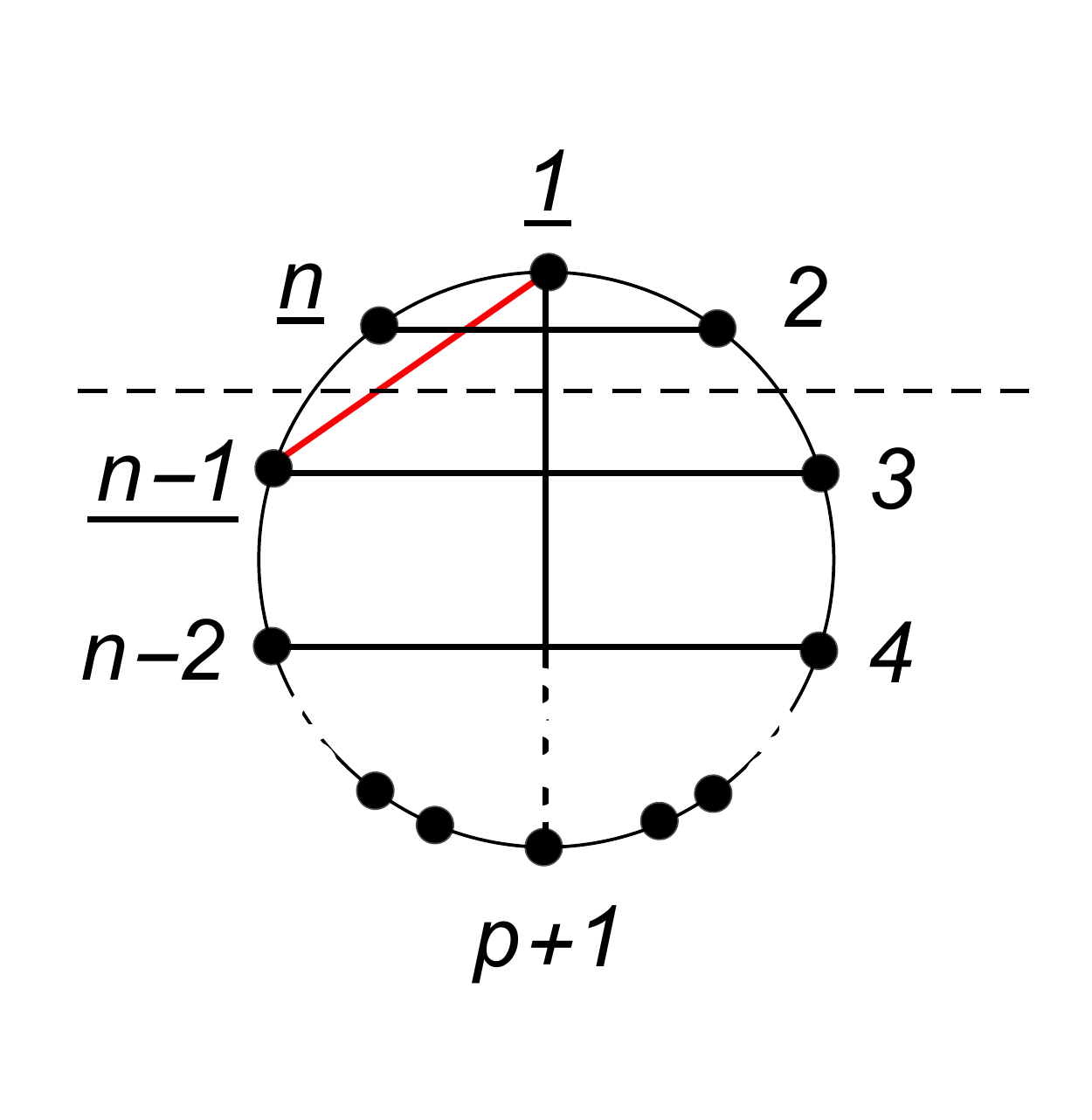}}
(b)
\hspace{-0.6cm}
\parbox[c]{10.2em}{\includegraphics[scale=0.23]{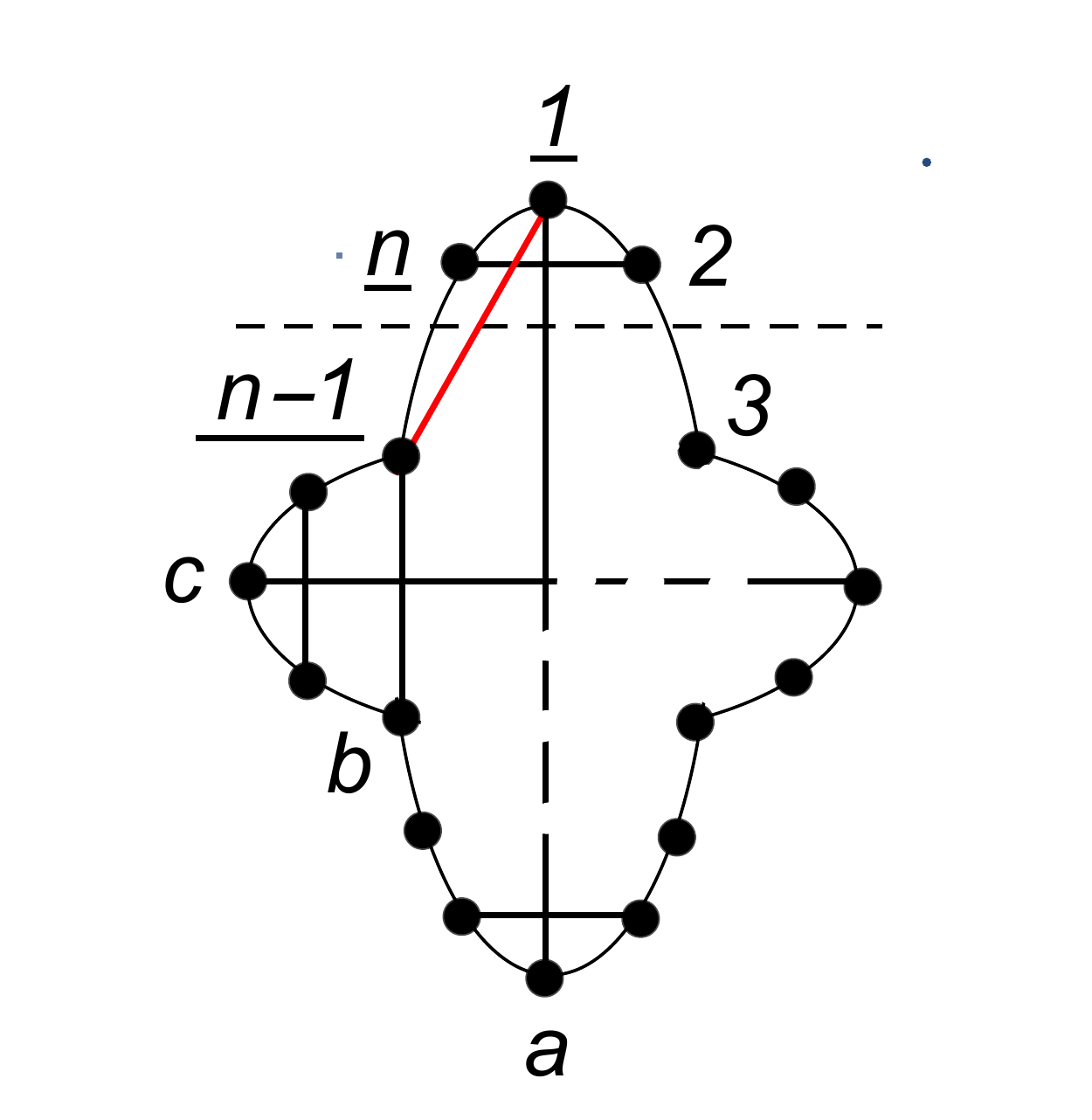}} 
\vspace{-0.2cm}
\caption{Factorization of (a) a ladder diagram, (b) a non-ladder diagram.}\label{Fig6} 
\vspace{-0.2cm}
\end{figure}

Let us briefly comment on the $n$-point computation. 
First notice that the six-point results can be straightforwardly extended to ladder diagrams with any number of points. In particular, let us consider the ladder diagram in Fig. \ref{Fig6}(a), where we fix the positions of legs $\left\{n-1,n,1\right\} $ and remove rows/columns $\left\{ 1,n-1\right\} $ from the $A$-matrix in the reduced Pfaffian. Like in the six-point case, only one factorization contributes, notably $\sigma_3\rightarrow\sigma_4 \rightarrow \cdots \rightarrow\sigma_{n-1}$.
Using the parametrization,  $\sigma_a=\epsilon x_a+\sigma_{n-1}$,  
with $a=3,4,\ldots,n-1$, $x_{n-2}=\textrm{constant}$, $x_{n-1}=0$, $\sigma_{n-1}\equiv\sigma_L$, and expanding around $\epsilon=0$, one obtains a generalization of  \eqref{eq:14}:  
\begin{align}\label{eq:18}
&
{\cal A}(\mathbb{1}_n \!: \! 1  (p+1),2n,... , p (p+2)) \,\, {\cal C}^\Delta_n
=\nonumber  \\
&
 {\cal A}(n,1,2,L:1L,2 n) \,\,
[( {\cal D}_{n}+{\cal D}_{1}+{\cal D}_{2})^2+m^2 ]^{-1} \, \nonumber\\
&
{\cal A}(R,3,..., n-1:R(p+1),..., p (p+2)) \,\, {\cal C}^\Delta_n ,
\end{align}
where  ${\cal A}(n,1,2,L:1L,2 n)$ is similar to \eqref{eq:15} and
\begin{multline}\label{eq:19}
{\cal A}(R,3,..., n-1:R(p+1),..., p (p+2)) = \\
\int_{\hat \gamma}  \prod_{a=3}^{n-3}  \dif x_a  \hat{S}^{-1}_a \frac{ {\rm Pf}A^{R(n-2)}_{R(n-2)}}{ x_{R(n-2)}}\\
\frac{
[x_{(n-2)(n-1)} x_{(n-1)R} x_{R(n-2)} ]^2   {\rm PT}(R,..., n-1)}{ (x_{R(p+1)}x_{3(n-1)}x_{4(n-2)} \cdots x_{p(p+2)}   )}, 
\end{multline}
with $x_R=\infty$. Here we have used the identity
\begin{equation}\label{pfaffianidentity}
\frac{ (-1) {\rm Pf}A^{R(n-1)}_{R(n-1)}}{ x_{R(n-1)}}  =  \frac{ {\rm Pf}A^{R(n-2)}_{R(n-2)}}{ x_{R(n-2)}}.
\end{equation} 
The integrand in \eqref{eq:19} reproduces the ladder diagram of Fig. \ref{Fig6}(a) with $(n-2)$ points, so equation \eqref{eq:18} provides a recursion relation. Since all terms in \eqref{eq:18} commute, this provides an inductive proof that we are free to shuffle terms in the Pfaffian with scattering equations, and there are no ambiguities in the definition of the integrand. 

Above six points, there are graphs with other topologies as depicted in Fig.\ref{Fig6}(b). A similar procedure can be used to build up such graphs by attaching 4-point vertices to diagrams with general topology, but there are additional complications because the Pfaffian identity in \eqref{pfaffianidentity} no longer applies \cite{W.in.P}.


\section{Discussion}

We have proposed a worldsheet description for cosmological correlators describing massive $\phi^4$ theory in de Sitter space. The scattering equations are written in terms of conformal generators which take a simple form in momentum space and make the flat space limit completely transparent. Another highly nontrivial ingredient of our formula is a Pfaffian defined in terms of the conformal generators.

There are a number of future directions to explore. In flat space, the scattering equations revolutionised the study of scattering amplitudes, revealing new perturbative dualities \cite{Cachazo:2014xea} and providing new tools for computing loop amplitudes \cite{Geyer:2015bja,Gomez:2017lhy,Farrow:2020voh} and soft limits \cite{Schwab:2014xua,Adamo:2014yya,Geyer:2014lca,Chakrabarti:2017ltl,Nandan:2016ohb}. We therefore expect the approach developed here will lead to similar progress for cosmological correlators. Moreover, although we have focused on the simplest toy model, there are systematic ways to make it more realistic such as using more general mass deformations in order to allow fields of different masses to propagate \cite{Naculich:2014naa}, considering different integrands which encode more general interactions, and breaking conformal symmetry since cosmological surveys measure correlators of curvature perturbations which become nontrivial when de Sitter boosts are broken \cite{Cheung:2007st,Green:2020ebl,Pajer:2020wxk}. It would also be interesting to extend our formula to spinning correlators. In this case, the external polarizations introduce a bigger class of operators than the operatorial Pfaffian we consider in this paper. Moreover, it would be of great interest to directly compute correlators by diagonalising the cosmological scattering equations. While this is analytically challenging, in practice it may be numerically feasible. Our momentum space formulae may be well-suited to this purpose given their simplicity compared to those in position space.


More ambitiously, we would like to investigate how to lift of our formula to that of a UV complete theory. As a first step, we can replace the scattering equations with Koba-Nielsen factors, and Mandelstam variables with differential operators in momentum space. This may lead to ordering ambiguities similar to those encountered when lifting the Virasoro-Shapiro amplitude to AdS$_5 \times$ S$^5$ \cite{Abl:2020dbx,Aprile:2020mus}. Ultimately, we hope that our worldsheet formula will provide a useful toy model for understanding the physics of the early Universe. 

\begin{acknowledgments}
We thank Sadra Jazayeri, Paul McFadden, Enrico Pajer, David Skinner, and David Stefanyszyn for useful conversations. HG and AL are supported by the Royal Society via a PDRA grant and a University Research
Fellowship, respectively. RLJ acknowledges the Czech Science Foundation - GA\v{C}R for financial support under the grant 19-06342Y.
\end{acknowledgments}

\end{document}